\documentclass[aps,prl,floatfix,twocolumn,showpacs]{revtex4}%
\usepackage{amsfonts}
\usepackage{amsmath}
\usepackage{amssymb}
\usepackage{graphicx}%
\begin{document}
\title{Positive cross-correlations in a three-terminal quantum dot with ferromagnetic contacts}
\author{A. Cottet, W. Belzig, and C. Bruder}
\affiliation{Department of Physics and Astronomy, University of Basel, Klingelbergstrasse
82, 4056 Basel, Switzerland}

\begin{abstract}
We study current fluctuations in an interacting three-terminal quantum dot
with ferromagnetic leads. For appropriately polarized contacts, the transport
through the dot is governed by a novel dynamical spin blockade, i.e., a
spin-dependent bunching of tunneling events not present in the paramagnetic
case. This leads for instance to positive zero-frequency cross-correlations of
the currents in the output leads even in the absence of spin accumulation on
the dot. We include the influence of spin-flip scattering and identify
favorable conditions for the experimental observation of this effect with
respect to polarization of the contacts and tunneling rates.

\end{abstract}
\date{\today}
\maketitle


Quantum fluctuations of current in mesoscopic devices have attracted
considerable attention in the last years (for reviews, see
Refs.~\cite{blanter:00,nazarov:03}). It has been shown that the statistics of
non-interacting fermions leads to a suppression of noise below the classical
Poisson value \cite{khlus:87,lesovik:89,buettiker:90} and to negative
cross-correlations in multi-terminal structures \cite{buettiker:92}. This was
recently confirmed experimentally in a Hanbury Brown-Twiss setup \cite{ccexp}.
The question of the sign of cross-correlations has triggered a lot of activity
\cite{buettiker:03-book}, and different mechanisms to obtain \textit{positive}
cross-correlations in electronic systems have been proposed. Employing a
superconductor as a source, positive cross-correlations have been predicted
for several setups \cite{positivecc}. This is because a superconducting source
injects highly correlated electron pairs. Screening currents due to long-range
Coulomb interactions lead to positive correlations in the finite-frequency
voltage noise measured at two capacitors coupled to a coherent conductor
\cite{martin:00,buettiker:03-book}. Lastly, positive cross-correlations can
occur due to the correlated injection of electrons by a voltage probe
\cite{texier:00}, or due to correlated excitations in a Luttinger liquid
\cite{safi:01}.

Below we will be interested in noise correlations in a quantum dot. This
problem was addressed theoretically in the sequential-tunneling limit
\cite{setnoise} and in the cotunneling regime \cite{setcotunneling}. Noise
measurements \cite{birk:95} were in agreement with the Coulomb-blockade
picture \cite{setnoise}. Cross-correlations between particle currents in a
paramagnetic multi-terminal quantum dot were studied in Ref.~\cite{bagrets:02}%
, and they were found to be negative. The noise of a two-terminal quantum dot
with ferromagnetic contacts was studied in the sequential tunneling limit
\cite{bulka:99,bulka:00}, and, interestingly, a super-Poissonian Fano factor
was found.

In this Letter, we consider an interacting three-terminal quantum dot with
ferromagnetic leads. The dot is operated as a beam splitter: one contact acts
as source and the other two as drains. Our main finding is that sufficiently
polarized contacts can lead to a dynamical spin blockade on the dot, i.e., a
spin-dependent bunching of tunneling events not present in the paramagnetic
case. A striking consequence of this spin blockade is the possibility of
positive cross-correlations in the absence of correlated injection.
Surprisingly, spin accumulation on the dot is not necessary to observe this
effect. Furthermore, the sign of cross-correlations can be switched by
reversing the magnetization of one contact. The effect is robust against
spin-flips on the dot as long as the spin-flip scattering rate is less than
the tunneling rates.

\begin{figure}[ptb]
\includegraphics[width=0.9\linewidth]{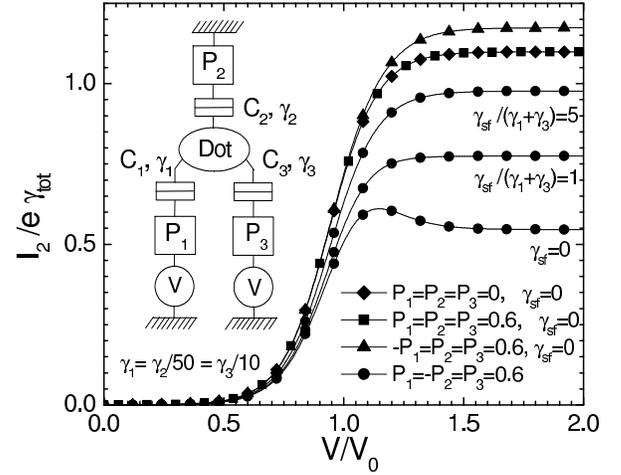}\caption{Current-voltage
characteristic of a quantum dot connected to three ferromagnetic leads
$i\in\{1,2,3\}$, with respective polarizations $P_{i}$, through tunnel
junctions with capacitances $C_{i}$ and net tunneling rates $\gamma_{i}$
(circuit shown in the inset). A voltage bias $V$ is applied to leads $1$ and
$3$; lead $2$ is connected to ground. The average current $I_{2}$ through lead
$2$ is shown as a function of voltage, for $C_{1}=C_{2}=C_{3}$, $\gamma
_{1}=\gamma_{2}/50=\gamma_{3}/10$, $k_{B}T/E_{0}=0.1$, and different values of
lead polarizations. The current is plotted in units of $e\gamma_{tot}%
=e\gamma_{2}(\gamma_{1}+\gamma_{3})/(\gamma_{1}+\gamma_{2}+\gamma_{3})$; the
voltage in units of $V_{0}=E_{0}C/(C_{1}+C_{3})e$; $E_{0}$ is the position of
the dot level. For $P_{1}=P_{2}=P_{3}$, $I_{2}$ coincides with the
paramagnetic case (diamonds). In the other cases, the high-voltage limit of
$I_{2}$ can be larger or smaller than the paramagnetic value, depending on the
lead polarizations. For $P_{1}=-P_{2}=P_{3}=0.6$ (circles), the effect of
spin-flip scattering is shown. Spin-flip scattering makes the $I_{2}-V$ curve
tend to the paramagnetic one.}%
\label{Figure1}%
\end{figure}

The system we have in mind is a quantum dot connected to three ferromagnetic
leads $i\in\{1,2,3\}$, through tunnel junctions with capacitances $C_{i}$ and
net spin-independent tunneling rates $\gamma_{i}$ (inset of Fig.~\ref{Figure1}%
). A voltage bias $V$ is applied to leads $1$ and $3$; lead $2$ is connected
to ground. At voltages and temperatures much lower than the intrinsic level
spacing and the charging energy $E_{C}=e^{2}/2C$ of the dot ($C=\sum_{i}C_{i}%
$), only one energy level of the dot located at $E_{0}$ needs to be taken into
account. In this situation, the dot can either be empty, or occupied with one
electron with spin $\sigma\in\{\uparrow,\downarrow\}$. Here and in the
following, we will measure energies from the Fermi level, i.e., $E_{F}=0$.

The collinear magnetic polarizations $P_{j}$ of the leads are taken into
account by using \textit{spin-dependent} tunneling rates $\gamma_{j\sigma
}=\gamma_{j}(1+\sigma P_{j})$, where $\sigma=\pm1$ labels the electron spin.
In a simple model, the spin-dependence is a consequence of the different
densities of states for majority and minority electrons \cite{julliere:75}.
The rate for an electron to tunnel on/off the dot ($\epsilon=\pm1$) through
junction $j$ is then given by $\Gamma_{j\sigma}^{\epsilon}=\gamma_{j\sigma
}/(1+\exp[\epsilon(E_{0}-eV_{j})/k_{B}T])$, where $V_{1}=V_{3}=-C_{2}V/C$,
$V_{2}=(C_{1}+C_{3})V/C$. On the dot, there can be spin-flip scattering, for
instance due to spin-orbit coupling or magnetic impurities. Here, we will
assume that the on-site energy on the dot does not depend on spin. Hence, due
to the detailed-balance rule, the spin-flip scattering rate $\gamma_{sf}$ does
not depend on spin.

In the sequential-tunneling limit $\hbar\gamma_{j\sigma}\ll k_{B}T$,
electronic transport through the dot can be described by the master equation
\cite{setnoise,bulka:00}:
\begin{equation}
\frac{d}{dt}\left[
\begin{array}
[c]{c}%
p_{\uparrow}(t)\\
p_{\downarrow}(t)\\
p_{0}(t)
\end{array}
\right]  =\hat{M}\left[
\begin{array}
[c]{c}%
p_{\uparrow}(t)\\
p_{\downarrow}(t)\\
p_{0}(t)
\end{array}
\right]  \;, \label{MasterEquation}%
\end{equation}
where $p_{\psi}(t)$, $\psi\in\{\uparrow,\downarrow,0\}$, is the instantaneous
occupation probability of state $\psi$ at time $t$, and where
\begin{equation}
\hat{M}=\left[
\begin{array}
[c]{ccc}%
-\Gamma_{\uparrow}^{-}-\gamma_{sf} & \gamma_{sf} & \Gamma_{\uparrow}^{+}\\
\gamma_{sf} & -\Gamma_{\downarrow}^{-}-\gamma_{sf} & \Gamma_{\downarrow}^{+}\\
\Gamma_{\uparrow}^{-} & \Gamma_{\downarrow}^{-} & -\Gamma_{\uparrow}%
^{+}-\Gamma_{\downarrow}^{+}%
\end{array}
\right]  \label{MatrixM}%
\end{equation}
depends on the total rates $\Gamma_{\sigma}^{\epsilon}=\sum_{j}\Gamma
_{j\sigma}^{\epsilon}$ and $\gamma_{\sigma}=\sum_{j}\gamma_{j\sigma}$. The
stationary occupation probabilities $\bar{p}_{\psi}$ are
\begin{equation}
\bar{p}_{\sigma}=\frac{\Gamma_{\sigma}^{+}\Gamma_{-\sigma}^{-}+\gamma
_{sf}(\Gamma_{\sigma}^{+}+\Gamma_{-\sigma}^{+})}{\gamma_{\sigma}%
\gamma_{-\sigma}-\Gamma_{\sigma}^{+}\Gamma_{-\sigma}^{+}+\gamma_{sf}%
(\Gamma_{\sigma}^{+}+\Gamma_{-\sigma}^{+}+\gamma_{\sigma}+\gamma_{-\sigma}%
)}\;,
\end{equation}
and $\bar{p}_{0}=1-\bar{p}_{\uparrow}-\bar{p}_{\downarrow}$. They can be used
to calculate the average value $\langle I_{j}\rangle$ of the tunneling current
$I_{j}(t)$ through junction $j$ as $\langle I_{j}\rangle=e\sum_{\epsilon
,\sigma}\epsilon\Gamma_{j\sigma}^{\epsilon}\bar{p}_{A(\sigma,-\epsilon)}$,
where $A(\sigma,\epsilon)$ is the state of the dot after the tunneling of an
electron with spin $\sigma$ in the direction $\epsilon$, i.e., $A(\sigma
,-1)=0$, $A(\sigma,+1)=\sigma$.

In the following, we first consider the situation $E_{0}>0$. The voltage $V$
will always be assumed to be positive, such that it is energetically more
favorable for electrons to go from the input electrode 2 to the output
electrodes 1 or 3 than in the opposite direction. The typical voltage
dependence of $I_{2}\equiv\langle I_{2}\rangle$ is shown in Fig.~\ref{Figure1}%
. The total current $I_{2}$ is exponentially suppressed at low voltages,
increases around a voltage $V_{0}=E_{0}C/(C_{1}+C_{3})e$, and saturates at
higher voltages. The width of the increase is determined by $k_{B}T/e$. The
high-voltage limit of $I_{2}$ depends on the polarizations $P_{i}$ and rates
$\gamma_{i}$ but not on the capacitances $C_{i}$. For a sample with magnetic
contacts, this limit can be higher or lower than that of the paramagnetic
case, depending on the parameters considered. In the high-voltage limit,
$I_{2}(P_{1},P_{2},P_{3})-I_{2}(0,0,0)=2e\gamma_{c}P_{out}\langle S\rangle$,
where $P_{out}=(P_{1}\gamma_{1}+P_{3}\gamma_{3})/(\gamma_{1}+\gamma_{3})$ is
the net output lead polarization, $\langle S\rangle=\nu(P_{2}-P_{out})$ is the
average spin accumulation on the dot \cite{Accumulation} and $\gamma
_{c}=\gamma_{2}(\gamma_{1}+\gamma_{3})/(\gamma_{1}+2\gamma_{2}+\gamma_{3})$.
Here, $\nu$ is a positive function of the polarizations, the tunneling and
scattering rates, which tends to $0$ at large $\gamma_{sf}$. Having a
saturation current different from the paramagnetic case requires $P_{out}%
\neq0$ \textit{and} $\langle S\rangle\neq0$. Spin-flip scattering modifies the
$I_{2}-V$ curve once $\gamma_{sf}$ is of the order of the tunneling rates. It
suppresses spin accumulation and makes the $I_{2}-V$ curve tend to the
paramagnetic one.

\begin{figure}[ptb]
\includegraphics[width=0.9\linewidth]{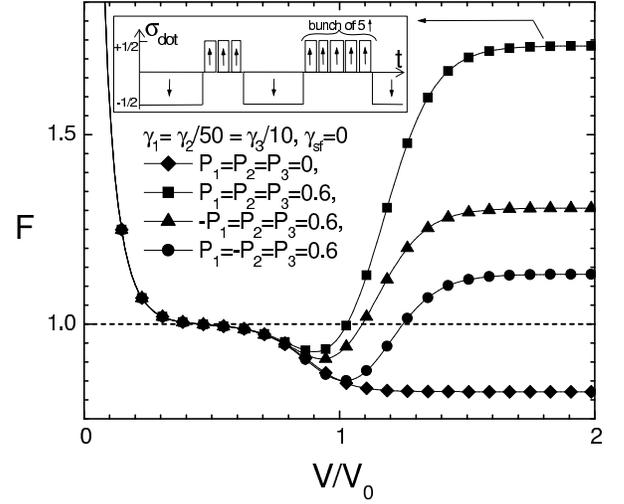}\caption{Fano factor
$F=S_{22}/2eI_{2}$ of lead 2 as a function of voltage, for the same circuit
parameters as in Fig. 1. In all curves $\gamma_{sf}=0.$ For $P_{1}=P_{2}%
=P_{3}$, the Fano factor is different from that of the paramagnetic case
(diamonds) in contrast to what happens for the average currents. The inset
shows the typical time dependence of the spin on the dot, in the high-voltage
limit $V\gg V_{0}$ for the case $P_{1}=P_{2}=P_{3}=0.6$. }%
\label{Figure2}%
\end{figure}

The power spectrum of tunneling current correlations in leads $i$ and $j$ is
defined as
\begin{equation}
S_{ij}(\omega)=2\int_{-\infty}^{+\infty}dt\exp(i\omega t)\langle\Delta
I_{i}(t)\Delta I_{j}(0)\rangle\,, \label{CorrelationsDefinition}%
\end{equation}
where $\Delta I_{i}(t)=I_{i}(t)-\langle I_{i}\rangle$. The terms $\langle
I_{i}(t)I_{j}(0)\rangle$ can be written as a function of the conditional
probabilities $P_{\psi,\varphi}^{c}(t)$ which are the occupation probabilities
of the state $\psi$ at time $t$ if at $t=0$ the state was $\varphi$, and which
are zero for $t<0$. Solving Eq.~(\ref{MasterEquation}) with the initial
condition $P_{\psi,\varphi}^{c}(t=0)=\delta_{\psi,\varphi}$ leads to
$P_{\psi,\varphi}^{c}(t)$. Its Fourier transform is given by $\hat{P^{c}%
}(\omega)=\int_{0}^{\infty}dt\exp(i\omega t)\hat{P^{c}}(t)=-(i\omega+\hat
{M})^{-1}$. The eigenvalues of the matrix $\hat{M}$ thus govern the frequency
dependence of $\hat{P}^{c}(\omega)$. The non-zero eigenvalues are
$\lambda_{\pm}=\frac{1}{2}\left(  -2\gamma_{sf}-\gamma_{\uparrow}%
-\gamma_{\downarrow}\pm\Delta\right)  $, with $\Delta^{2}=4\gamma_{sf}%
^{2}+(\gamma_{\uparrow}-\gamma_{\downarrow})^{2}-4\gamma_{sf}(\Gamma
_{\uparrow}^{+}+\Gamma_{\downarrow}^{+})+4\Gamma_{\uparrow}^{+}\Gamma
_{\downarrow}^{+}$. This eventually leads to
\begin{equation}
S_{ij}(\omega)=\delta_{ij}S_{j}^{Sch}+\sum_{\sigma,\sigma^{\prime}}%
S_{i,\sigma,j,\sigma^{\prime}}^{c}(\omega)\;, \label{eq:noise1}%
\end{equation}
where $S_{j}^{Sch}=2e^{2}\sum_{\epsilon,\sigma}\Gamma_{j\sigma}^{\epsilon}%
\bar{p}_{A(-\epsilon,\sigma)}$ is the Schottky noise produced by tunneling
through junction $j$, and
\begin{align}
\frac{S_{i,\sigma,j,\sigma^{\prime}}^{c}(\omega)}{2e^{2}}  &  =\sum
_{\epsilon,\epsilon^{\prime}}\epsilon\epsilon^{\prime}\left[  \Gamma_{i\sigma
}^{\epsilon^{\prime}}G_{A(\sigma,-\epsilon^{\prime}),A(\sigma^{\prime
},\epsilon)}(\omega)\Gamma_{j\sigma^{\prime}}^{\epsilon}\bar{p}_{A(\sigma
^{\prime},-\epsilon)}\right. \label{eq: noise2}\\
&  \left.  +\Gamma_{{j}\sigma^{\prime}}^{\epsilon^{\prime}}G_{A(\sigma
^{\prime},-\epsilon^{\prime}),A(\sigma,\epsilon)}(-\omega)\Gamma_{i\sigma
}^{\epsilon}\bar{p}_{A(\sigma,-\epsilon)}\right]  \,.\nonumber
\end{align}
Here, we defined $G_{\psi,\varphi}(\omega)=P_{\psi,\varphi}^{c}(\omega
)+\bar{p}_{\psi}/i\omega.$

Equation~(\ref{eq:noise1}) determines the full frequency-dependent tunneling
current correlation functions of the three-terminal quantum dot. For
frequencies larger than the cutoff frequencies $\lambda_{-}$ and $\lambda_{+}
$, the spectrum $S_{ij}(\omega)$ tends to the uncorrelated spectrum
$\delta_{ij}S_{j}^{Sch}$. In the following, we will consider mainly the
zero-frequency limit of $S_{ij}(\omega)$, because the frequencies
$\lambda_{\pm}\sim\gamma_{i}$ are difficult to access in experiment. Note that
at zero frequency, the contribution of the screening currents ensuring
electro-neutrality of the capacitors after a tunneling event
\cite{buettiker:03-book} is zero, i.e., $S_{ij}\equiv S_{ij}(0)$ is the signal
measured in practice \cite{screening}.

\begin{figure}[tb]
\includegraphics[width=0.9\linewidth]{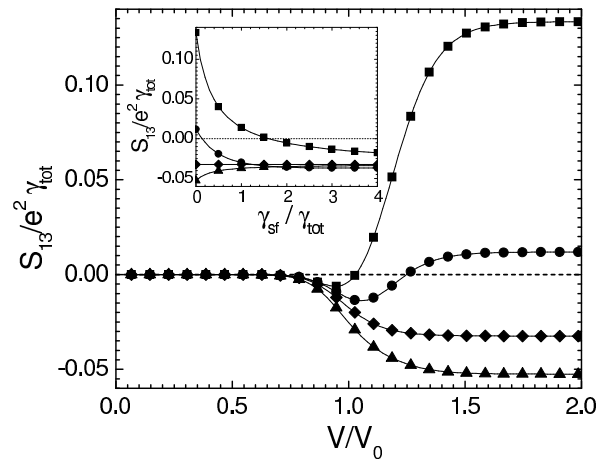}\caption{Current
cross-correlations between leads $1$ and $3$ as a function of voltage. The
curves are shown for the same circuit parameters as in Fig.~\ref{Figure2}. The
cross-correlations can be positive in the cases $P_{1}=-P_{2}=P_{3}=0.6$
(circles) and $P_{1}=P_{2}=P_{3}=0.6$ (squares). Note that the sign of
cross-correlations can be reversed just changing the sign of $P_{1}$. In all
curves $\gamma_{sf}=0$. The inset shows the influence of spin-flip scattering
on the cross-correlations in the high-voltage limit $V\gg V_{0}$. In the
paramagnetic case (diamonds), spin-flip scattering has no effect. In the limit
$\gamma_{sf}\gg\gamma_{tot}$, the cross-correlations tend to the paramagnetic
value. }%
\label{Figure3}%
\end{figure}

Figures~\ref{Figure2}, \ref{Figure3} show the Fano factor $F=S_{22}/2eI_{2}$
and the cross-correlations $S_{13}$ as a function of $V$ for $\gamma_{sf}=0$.
Well below $V_{0}$ the current is due to thermally activated tunneling and the
noise is Poissonian. At very low voltage, $eV\leq k_{B}T$, the cross-over to
thermal noise is observed. Around $V=V_{0}$, $F$ and $S_{13}$ show a step or a
dip. The high-voltage limit strongly depends on tunneling rates and
polarizations. In the paramagnetic case, the limit of $F$ lies in the interval
$[1/2,1]$, and that of $S_{13}/2eI_{2}$ in $[-1/8,0]$. In the ferromagnetic
case the high-voltage limit of $F$ can be either sub- or super-Poissonian, as
already pointed out in the two-terminal case \cite{bulka:99}. Spin
accumulation is not a necessary condition for having a super-Poissonian Fano
factor, as can be seen for $P_{1}=P_{2}=P_{3}$, where $\left\langle
S\right\rangle =0.$ In this case, the essential point is that the current can
flow only in short time windows where the current transport is not blocked by
a down spin, see the inset of Fig. \ref{Figure2}. This dynamical spin blockade
leads to a bunching of tunneling events, and explains the super-Poissonian
Fano factor.

The cross-correlations can be either positive or negative, see
Fig.~\ref{Figure3}. Note that a super-Poissonian $F$ does not necessarily
imply positive cross-correlations, as shown by the case $-P_{1}=P_{2}%
=P_{3}=0.6$ in Figs.~\ref{Figure2}, \ref{Figure3}, for which the
cross-correlations are even more negative than in the paramagnetic case.
Indeed, relations (\ref{eq:noise1}) and (\ref{eq: noise2}) together with
charge conservation imply that $S_{22}-S_{2}^{Sch}=\sum_{\sigma,\sigma
^{\prime}}S_{1,\sigma,3,\sigma^{\prime}}^{c}\left(  \gamma_{1\sigma}%
+\gamma_{3\sigma}\right)  \left(  \gamma_{1\sigma^{\prime}}+\gamma
_{3\sigma^{\prime}}\right)  /\gamma_{1\sigma}\gamma_{3\sigma^{\prime}}$ at $V
\gg V_{0}$. Thus, at $V\gg V_{0}$, a super-Poissonian $F$ is equivalent to
positive cross-correlations only at large voltages and if the two output leads
have identical polarisations. For the case $-P_{1}=P_{2}=P_{3}=0.6$,
cross-correlations are negative in spite of the super-Poissonian $F$ because
the correlated electrons are mostly up electrons flowing through lead 3. We
note here that $\text{Re}[S_{13}(\omega)]$ can change sign for intermediate
frequencies and vanishes for $\omega\gg\lambda_{+},\lambda_{-}$
\cite{unpublished}.

We now briefly comment on the case $E_{0}<0$. For $V\gg V_{0}^{\prime}= -E_{0}
C/C_{2}e$, the values of $I_{2}$, $F$ and $S_{13}$ are the same as previously.
When $V$ is smaller than $V_{0}^{\prime}$, the most striking difference is
that $F$ is polarization-dependent and thus not necessarily Poissonian.

The effect of spin-flip scattering is shown in the inset of Fig.~\ref{Figure3}%
. Spin-flip scattering influences the cross-correlations once $\gamma_{sf}$ is
of the order of the tunneling rates. In the high-$\gamma_{sf} $ limit,
cross-correlations tend to the paramagnetic case for any value of the
polarizations. Thus, strong elastic spin-flip scattering suppresses positive
cross-correlations, in contrast to what happens with inelastic scattering in
\cite{texier:00}. In practice, experiments with a quantum dot connected to
ferromagnetic leads and $\gamma_{sf}\ll\gamma_{tot}$ have already been
performed \cite{Deshmukh}. Thus, spin-flip scattering should not prevent the
observation of positive cross-correlations in quantum dots.

\begin{figure}[tb]
\includegraphics[width=0.9\linewidth]{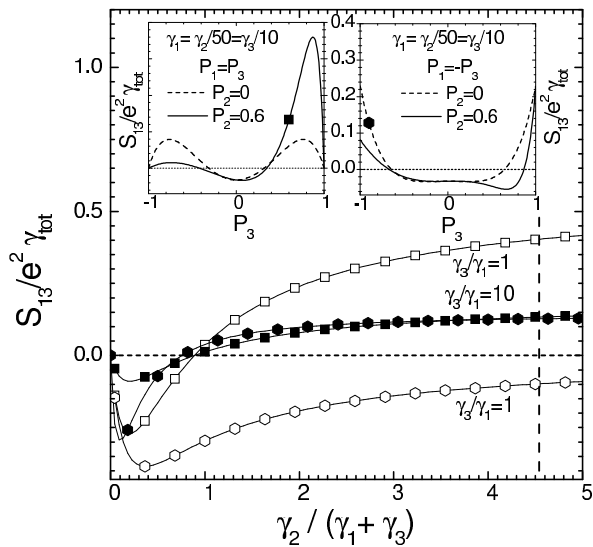}\caption{Influence of
asymmetry between $\gamma_{2}$ and $\gamma_{1}+\gamma_{3}$ on the high-voltage
limit of the cross-correlations, for $P_{1}=P_{3}=P_{3}=0.6$ (squares) and
$-P_{1}=P_{3}=0.9$, $P_{2}=0$ (hexagons), for $\gamma_{3}/\gamma_{1}=10$ (full
symbols) and $\gamma_{3}/\gamma_{1}=1 $ (empty symbols). Large values of
$\gamma_{2}/(\gamma_{1}+\gamma_{3})$ favor positive cross-correlations. For
$-P_{1}=P_{3}=0.9 $, $P_{2}=0$, an asymmetry between $\gamma_{1}$ and
$\gamma_{3}$ is also necessary. The vertical dashed line indicates the ratio
$\gamma_{2}/(\gamma_{1}+\gamma_{3})$ corresponding to Figs.~\ref{Figure1},
\ref{Figure2}. The two insets show the high-voltage limit of the
cross-correlations as a function of $P_{3}$, for $\gamma_{1}=\gamma
_{2}/50=\gamma_{3}/10$, $P_{1}=P_{3}$ (left inset) and $P_{1}=-P_{3}$ (right
inset) and $P_{2}=0$ (dashed lines) or $P_{2}=0.6$ (full lines). For all
curves $\gamma_{sf}=0$.}%
\label{Figure4}%
\end{figure}

Finally, we address the problem of how to choose parameters that favor the
observation of positive cross-correlations. First, finite lead polarizations
are necessary \cite{bagrets:02}, see the insets of Fig.~\ref{Figure4}.
However, it is possible to get positive cross-correlations even if $P_{2}=0$,
provided the output leads 1,3 of the device are sufficiently polarized (dashed
lines in the insets of Fig.~\ref{Figure4}). The case where the three
electrodes are polarized in the same direction seems the most favorable. In
the high-voltage limit, choosing $P_{1}=P_{2}=P_{3}$ and $\gamma_{sf}=0$ leads
to
\begin{equation}
S_{13}=\frac{16e^{2}\gamma_{1}\gamma_{2}^{2}\gamma_{3}[(\gamma_{1}+2\gamma
_{2}+\gamma_{3})P_{1}^{2}-\gamma_{1}-\gamma_{3}]}{(\gamma_{1}+\gamma
_{3})(\gamma_{1}+2\gamma_{2}+\gamma_{3})^{3}(1-P_{1}^{2})}\;.
\label{Sexplicit}%
\end{equation}
The asymmetry between the tunneling rates $\gamma_{i}$ has a strong influence
on the cross-correlations, see Fig.~\ref{Figure4}. Large values of $\gamma
_{2}/(\gamma_{1}+\gamma_{3})$ favor the observation of positive
cross-correlations, see e.~g.~Eq.~(\ref{Sexplicit}), by decreasing $\bar
{p}_{0}$. This allows to extend the domains of positive cross-correlations to
smaller values of the polarizations, which is important because experimental
contact materials are not fully polarized. For $\gamma_{1}=\gamma
_{2}/10=\gamma_{3}$, the polarizations $P_{1}=P_{2}=P_{3}=0.4$ typical for Co
\cite{Soulen} lead to positive cross-correlations of the order of
$S_{13}/e^{2}\gamma_{tot}\simeq0.08$. With $\gamma_{tot}\simeq5$GHz this
corresponds to $10^{-29}$A$^{2}$s, a noise level accessible with present
noise-amplification techniques \cite{birk:95}.

In conclusion, we have demonstrated that transport through a multi-terminal
quantum dot with ferromagnetic contacts is characterized by a new mechanism,
viz., dynamical spin blockade. As one of its consequences we predict positive
current cross-correlations in the drain contacts without requiring the
injection of correlated electron pairs. We have included spin-flip scattering
on the dot and have shown that the effect persists as long as the spin-flip
scattering rate is less than the tunneling rate to the contacts.

We thank T. Kontos and C. Sch\"{o}nenberger for useful discussions. This work
was financially supported by the RTN Spintronics, by the Swiss NSF, and the
NCCR Nanoscience.

\end{document}